\documentclass[preprint, 12pt]{elsarticle}

\usepackage{amssymb}
\usepackage{float}
\usepackage[colorlinks=true,
            linkcolor=blue,
            citecolor=blue,
            urlcolor=blue]{hyperref}

\usepackage{amsmath}
\usepackage{nomencl}
\usepackage{xcolor}
\usepackage[utf8]{inputenc}

\journal{Applied Energy}

\begin{document}

\begin{frontmatter}

\title{Assessing the Shortfall Risk of GB Electricity Grid using Shifts in Winter Weather Conditions}

 \author[som1]{Aninda Bhattacharya}
 \author[som1]{Chris J. Dent}
 \author[som1]{Amy L. Wilson}
 \author[som2]{Gabriele C. Hegerl}
 
 \affiliation[som1]{organization={School of Mathematics and Maxwell Institute for Mathematical Sciences},
             addressline={University of Edinburgh},
             city={Edinburgh},
             postcode={EH9 3FD},
             country={United Kingdom}}

 \affiliation[som2]{organization={School of GeoSciences},
             addressline={University of Edinburgh},
             city={Edinburgh},
             postcode={EH8 9XP},
             country={United Kingdom}}

\begin{abstract}
Extreme weather events during peak winter periods drive resource adequacy risk in Great Britain (GB), with weather sensitivity of the supply demand balance increasing through additional electric heating and wind generation. This work develops an approach of time-shifting weather within the peak season, through adjustment of the relevant terms in a statistical model for demand. This allows more complete consideration of the security of supply consequences of a weather series, as there will be relevant conditions where demand is suppressed due to weather occurring at a weekend or during the Christmas holiday. Results on a GB example show that consideration of this counterfactual is indeed important, and specifically that winter 2010-11 can either be the most severe in the dataset, or insignificant within the resource adequacy model, depending on the alignment of day-of-week with the weather series. Statistical interpretation of the shift model is discussed, which is straightforward for alignment of day-of-week with weather assuming that all seven alignments are equiprobable; but is more subtle for shifting weather in and out of Christmas, as there is no natural maximum on the realistic length of shift, but too large a shift may  be physically unrealistic. It is likely that in all systems assessment of a weather year’s severity is incomplete without such consideration of the day-of-week effect, however whether longer shifts of weather with respect to date need to be considered also will depend on the presence of a major holiday (such as Christmas in GB) in the peak season.

\end{abstract}

\begin{highlights}
\item Weather sensitivity of the supply demand balance is increasing through Great Britain, through additional electric heating and wind generation.
\item An approach for shifting alignment of weather with day-of-week and date is developed, through adjustment of the relevant terms in a statistical model of demand. 
\item Results on a GB example demonstrate how without this, the effect of important weather days may be masked by weekends and holidays, and thus the potential impact of a weather time series has not been fully assessed.
\item Statistical interpretation of this result, as relevant to inclusion in risk assessments, is discussed.
\item It is likely that consideration of shifting alignment of day-of-week and weather is important in all systems, though not all systems will see suppression of demand by an extended holiday in the peak system, such as Christmas in GB. 
\end{highlights}


\end{frontmatter}

\section{Introduction}
\label{sec:Introduction}

Security of supply is one of the primary drivers of electricity system development, along with cost and environmental impact, in GB \cite{UK_Statutory_Security_2025} and other countries. Key developments in decarbonisation include the growth of renewable generation, and electrification of heating, which respectively make supply availability and demand more weather sensitive.

Resource adequacy (RA) studies the risk of there being insufficient resources (generation, flexible demands etc.) to meet demand, either on operational planning timescales of months to a year ahead, or on longer capital planning timescales. This has traditionally been measured by indices such as LOLE (Loss of Load Expectation),  the expected number of periods (usually days or hours) in which there is a shortfall \cite{LOLE1, Keane}, or  Expected Energy Unserved (EEU). Terminology differs between N America (where LOLE generally refers to the expected number of days on which there is a shortfall, and the hourly measure is referred to as LOLH), and Europe (where the hourly measure is called LOLE) \cite{LOLE1}. Given the increasing variability of inputs due to temperature sensitivity, there is interest in using a wider range of model outputs which capture the corresponding variability of system outcome about the expected value \cite{9586399, ResourceAdequacy1}.

The dependence of available renewable supply on weather is widely studied in resource adequacy calculations \cite{Keane, ESIG2021_resource_adequacy1}. In addition, a potential major challenge Britain  faces is increased electrification of heat to meet its net-zero targets \cite{DESNZ2025CleanPower2030}, 
as most heating is presently by natural gas. Independent decarbonisation scenarios typically project a large increase in electricity demand through increasing deployment of  heat pumps and electric vehicles \cite{Quiggin2016The, electricitydemand3, electricitydemand2}. This has been projected to increase residential electricity demand by 50 per cent in 2035, and to double peak electricity demand by the end of 2050 \cite{Eggimann2020How,electricitydemand3}. \cite{electricitydemand2} estimates that a heat pump penetration of around 41\% can almost double the present electricity demand, with a large increase in inter-annual variability. 
Here, this potential increase in electric heating will be accounted for through a scenario of high sensitivity of demand to temperature, following the general philosophy of \cite{lucie} in using  parametrisation of  system scenarios to give transparent sensitivity analysis.

Estimation of the above indices requires statistical estimation of the process of demand minus renewable supply, for the future scenario for which the risk calculation is performed. The most common approach to this is `hindcast’, which asks what the risk level would be conditional on a repeat of a historic weather year; a headline or unconditional risk index is then calculated as the mean across the results conditional on multiple different available weather years \cite{Keane, ResourceAdequacy1}. 
When the capacity of renewables or electric heating/cooling grows, shortfall risk becomes dominated by fewer years (and days within them), and the sampling uncertainty associated with estimates of risk indices grows \cite{lucie}. 

In the Great Britain (GB) system, demand is suppressed significantly at weekends and during the 2 week Christmas holiday, to an extent that if a system has an acceptable probability of shortfall on  winter weekdays outside Christmas, then in these periods the risk of shortfalls will be minimal. However, the historical weather that fell on weekends is just as relevant to estimation of risk as that which fell at weekdays; the same can be said of weather that occurred during Christmas, and which might also have occurred in mid December or early January. This paper therefore develops an approach to unmasking the consequences of  weather which occurs at Christmas or at weekends, by shifting the relative alignments of weather, day of week and date, while maintaining the sequence of weather from the historic year.

Electricity demand is a complex interplay of economic, temporal, societal and human behavioural factors. However to make the required shifts of day-of-week or weather relative to date, and to map historic demand or weather series to demand in a future scenario, we simply require a statistical model of the relationship between demand and weather.  There are numerous approaches to modelling demand that range from using statistical models such as autoregressive time series \cite{tdemandmodel1, tdemandmodel2, tdemandmodel3, tdemandmodel4} and  multiple linear regression \cite{rdemandmodel1, rdemandmodel2, rdemandmodel3, demandmodel3, demandmodel4, Hyndman_2010}, to more sophisticated deep learning frameworks, such as neural networks \cite{ndemandmodel1, ndemandmodel2, demandmodel5}, depending on the purpose \cite{demandmodel2}. 
There has been interest in use of time series \cite{DeSilva2025} and machine learning   \cite{HUANG2025137205, demandmodel6} frameworks for long-term  projections, however caution is required here as long-term trends depend on structural changes in technology, economic and societal factors,  which might not naturally be modelled through estimation from past data. The structures from regression models may, however, be important in mapping changes in these underlying factors to demand in future scenarios \cite{Bloomfield_2016, DEAKIN2021117261, Edwheatcroft, Amy1}, and this  approach  is taken here; machine learning models may lack the interpretability required for this purpose.

The primary contributions of the present paper are 
\begin{itemize}
    \item The approach to shifting the alignment of day-of-week with weather, through shifting either weather or day-of-week with respect to date to more fully sample weather influence on demand and supply
    \item Demonstration on a GB test example that, without  considering all possible alignments of weather with working week days, the resource adequacy consequences of a given weather year have not been fully  assessed;
    \item Discussion of the statistical interpretation of these shifts of alignment of day-of-week or date with weather. 
\end{itemize}
This is supported by careful validation of the regression model for demand, including consideration of whether there is evidence for a special effect on demand of the most extreme weather conditions beyond the general statistical model. 

The paper is structured as follows: \autoref{sec:RMS} describes the  risk model and the data used in this paper. \autoref{sec:MD} then introduces the approach to shifting day of week and weather relative to date. Results, including how the new approach can give a broader picture of the potential consequences of a given weather year, are presented in \autoref{sec:LOLE}, and the statistical interpretation of these results is discussed in \autoref{sec:SI of LOLE}. Finally, conclusions are presented in \autoref{sec:conc}.

\section{Risk Model Specification}
\label{sec:RMS}

Various metrics are used in resource adequacy assessments to assess shortfall risks in different regions \cite{ResourceAdequacy1}. Here, we use Loss of Load Expectation (LOLE),  the expected number of periods in which the supply is insufficient to meet demand \cite{LOLE1}. Since the demand model used is for GB daily peak demand (at 6 PM during the winter months), for most results we use the `daily' LOLE metric, i.e. the expected number of daily demand peaks per year at which there is a shortfall (similar to the common long-standing  practice in N America). We also demonstrate that analogous results are obtained using `hourly' LOLE, the expected number of hours of shortfall per year (called LOLH in N America). This section describes the risk model used, and the weather and demand data on which results are based.

\subsection{Weather and Demand Data}
\label{subsec:Weather and Demand Data}

\subsubsection{Weather Data}
  The meteorological data used in this study are derived from the ERA-5 reanalysis, as described in \cite{weather-data-1}. Population-weighted temperature and 10m wind speed values are used as inputs to the regression for daily peak demand. The population weightings in the temperature and wind values were obtained from \cite{Doxsey-Whitfield03072015}.
  
  For wind generation, we use the onshore and offshore capacity factor time series in \cite{thewindpowerWindEnergy}, which combine historic wind speeds with a scenario of wind farm locations. The database includes wind farm locations and installed capacity data for the present day (2021 here, `current' scenaio), and  future projections of wind farm locations in 2030 (`future' scenario). Unlike demand, mapping offshore and onshore capacity factors to a target scenario is already done in \cite{weather-data-1}. Mapping wind-speed to wind generation does not require consideration of different temporal cycles over the period of interest, as with generation, it is only dependent on wind speed and the farm operating capacity at a particular point in time.  
  

\subsubsection{Demand Data}
For demand, we have used an hourly, national-area-averaged time series of gross demand for the winters from 2009-10 to 2019-20, obtained from the National Grid Data portal \cite{demanddata1}. The national  demand here represents the sum of the transmission-metered demand with distribution-connected renewable generation added to the metered sum (so that all renewable generation is treated as generation in subsequent analysis), but excludes  station load, the pumping of hydro storage, and the exports via interconnectors \cite{demanddata1}. This will be used  in \autoref{sec:MD} to fit the model mapping weather conditions to demand.


\subsection{Conventional Generation} \label{subsec:CG}


Capacity and availability probability data for the conventional generators in GB were obtained from  National Grid, as in \cite{SANCHEZ2023101151}, and were anonymised using a Gaussian error term before use due to sensitivity of the raw data -- the resulting data are generally representative of the real GB system scenarios for the analysis presented here. The availability each of conventional unit is taken to be zero or maximum, and assuming independence between availabilities of different units distribution of  total available conventional capacity $X$ is constructed using the standard convolution approach.


\subsection{System Scenarios Studied}

Four system scenarios are considered here. 
Within our  model these are defined by the installed capacity of wind power (onshore and offshore), and the sensitivity of demand to temperature and wind speed ($\lambda_p$ and $\gamma_p$ respectively, representing the volume of electric heating). In addition to the certain large growth in wind energy, the potential large growth in electric heating \cite{Eggimann2020How} means that a range of scenarios for weather sensitivity of demand must be studied -- for more discussion of this, see \cite{lucie}.

\begin{table}[H]
\begin{center}
    \begin{tabular}{ccccc}
    \textbf{\begin{tabular}[c]{@{}c@{}}Target \\ Scenarios\end{tabular}}              & \textbf{\begin{tabular}[c]{@{}c@{}} $\mathbf{\lambda_p}$ \\ (GW/C)\end{tabular}} & \textbf{\begin{tabular}[c]{@{}c@{}}$\mathbf{\gamma_p}$\\ (GW / ms$^{-1}$)\end{tabular}} & \textbf{\begin{tabular}[c]{@{}c@{}} Wind Capacity\\ (GW)\end{tabular}} \\
    \hline
    \begin{tabular}[c]{@{}c@{}}Scenario 1\\ (Present Day)\end{tabular}                & -0.6                                                                                & 0.125                                                                                                                                            & \begin{tabular}[c]{@{}c@{}}Offshore = 16\\ Onshore = 14\end{tabular}               \\
    \hline
    \begin{tabular}[c]{@{}c@{}}Scenario 2\\ (Future wind, \\ Low heating)\end{tabular} & -0.6                                                                                & 0.125                                                                                                                                            & \begin{tabular}[c]{@{}c@{}}Offshore = 40\\ Onshore = 25\end{tabular}               \\
    \hline
    \begin{tabular}[c]{@{}c@{}}Scenario 3\\ (Future wind,\\ Medium heating)\end{tabular} & -1.2                                                                                & 0.25                                                                                                                                             & \begin{tabular}[c]{@{}c@{}}Offshore = 40\\ Onshore = 25\end{tabular}               \\
    \hline
    \begin{tabular}[c]{@{}c@{}}Scenario 4\\ (Future wind,\\ High heating)\end{tabular}     & -2.0                                                                                & 0.42                                                                                                                                             & \begin{tabular}[c]{@{}c@{}}Offshore = 40\\ Onshore = 25\end{tabular} 
    \end{tabular}
\end{center}
\caption{System scenarios considered in this study. Scenario 1 is representative of the present system, while Scenarios 2 to 4 are  representative of possible futures. Low/medium/high heating load is represented through temperature sensitivity of demand, and future systems with increased electric heating, respectively. 'Present' and 'future' wind refers to the installed capacity of wind generation.}
\label{tab:Scenarios}
\end{table}

\subsection{Hindcast Estimate of LOLE}

Once the demand and generation are mapped to a target scenario, the risk of shortfall in terms of LOLE is estimated. With available conventional capacity, available wind capacity and demand at time $t$ in the future year studied denoted $X_t$, $W_t$ and $D_t$, respectively, the surplus is  $Z_t = X_t - (D_t - W_t)$. Standard notation is used in which capital letters denote random variables and lower case denotes specific values. The Loss of Load Expectation  is then: 

\begin{equation}
    \text{LOLE} = \mathbb{E} \Biggl[\sum_{t } \mathbb{I }(Z_t < 0) \Biggr] =\sum_{t }  \mathbb{P } (Z_t < 0) ,
\label{HC1}
\end{equation}
where $\mathbb{I}$ is an indicator function taking value $1$ if the argument is true and $0$ otherwise.

The most common way of estimating LOLE based on historic data is so-called Hindcast, which can be interpreted either as using the empirical historical joint demand as the predictive distribution, or as the LOLE conditional on a repeat of historic weather conditions. The estimate conditional on an historic year $i$ is then 
\begin{equation}
    \text{LOLE$_i$} = \sum_{\tau \in T_i} \mathbb{P }(X_{\tau} < d_{\tau} - w_{\tau})
\label{HC2}
\end{equation}
where $\tau$ is used as the dummy index for historic times and $T_i$ is the set of times in year $i$. It is common to make an estimate of LOLE not conditional on a particular weather year, by averaging over the values conditional on the historic years $T_i$.

\section{Modelling Demand Using Shifts in Weather}
\label{sec:MD}

\subsection{Explanatory Weather Variables for Demand}

The temperature variable used for modelling demand is the `effective temperature', as specified by the National Energy System Operator for their demand modelling \cite{Edwheatcroft}. If the population-weighted temperature for hour $h$ in winter $i$ is $[TA]_{i,h}$, and $[TO]_{i,h}$ is the mean of the last four hours of $[TA]_{i,h}$, the effective temperature, $[TE]_{i,h}$ is defined as:
\begin{equation}
\begin{aligned}
    TE_{i,h} = \frac{1}{2}([TE]_{i,h-24} + [TO]_{i,h})
\end{aligned}
\label{TE1}
\end{equation}
where $[TE]_{i,h-24}$ is the effective temperature value 24 hours previously.  TE is designed to account for  daily and hourly lags in the effect of temperature on demand. Since we want to model the daily peak demand, only TE at 6pm each day will be used, denoted $[TE]_{i,t}$ for day $t$ in year $i$. As described earlier, population-weighted mean 10m wind speed (denoted $\gamma_1[WS]_{i,t}$) is used to model the wind chill effect on heat demand. 

\subsection{Linear Model for Daily Peak Demand}

Electricity demand is a complex interplay of societal, economic, and temporal factors beyond weather alone. The model used to explain daily peak demand, $\mathrm{D^{EMP}_{i,t}}$, thus includes, in addition to the weather variables $TE_{i,h}$ and, $[WS]_{i,t}$ described above, day-of-week, time-of-year, and year effects:
\begin{equation}
\begin{aligned}
    \mathrm{D^{EMP}_{i,t}} &= \alpha + \lambda_1 [TE]_{i,t}  + \beta_1[DSN]_t  + \beta_2[DSN]_t^2  \\
    &\quad    + \sum_{m=1}^6 \omega_m[DOW]_{m,i,t} + \gamma_1[WS]_{i,t}\\
    &\quad   + \sum^{2018}_{i=2009} \phi_i[Y]_i  + \epsilon_{i,t} \\
    &= \mathrm{\hat{D}_{i,t}} + \epsilon_{i,t}
\end{aligned}
\label{DM1}
\end{equation}
where: 
\begin{itemize}
    \item $\mathrm{D^{EMP}_{i,t}}$ is the daily peak demand at time $t$ in winter $i$.
    \item $[DSN]_t$ is the number of days since the start of November. The linear and quadratic terms in this variable capture seasonal effects on demand, principally due to lighting load. 
    \item $[Y]_i$ is an indicator variable taking value 1 for year $i$ and 0 otherwise, and its coefficient $\phi_i$ is the corresponding year effect (i.e. the underlying changes in what demands are connected to the system.) 2019 is taken as the reference year, and thus the corresponding indicator variable is excluded.
    \item $[DOW]_{m,i,t}$ is an indicator variable taking value 1 if date $(i,t)$ is day-of-week $m$, and 0 otherwise.
\end{itemize}
It is convenient for some of the analysis that follows to denote the central estimate of demand from the regression formula (i.e. the regression formula with the components explained by weather and time but without the residual term) as $\mathrm{\hat{D}_{i,t}}$.

\subsection{Assessing the fit of the daily peak demand model} \label{sec:assessfit}

To assess the fit of the regression model \eqref{DM1} to historical demand, several diagnostic checks have been performed. The purpose of performing this fit is to identify an appropriate set of variables for explaining historic demands, and to obtain baseline levels of these parameters from the recent history. It is particularly important to avoid a model with substantial overfitting or confounding effects, so that the coefficient values obtained are interpretable -- to this end, we found that the combination of individual year effects with constant linear wind speed and temperature effects gave the best overall outcome, and that allowing the weather sensitivities to vary between years gave minimal improvement in the quality of fit. Adding the wind speed term in addition to the temperature term did, however, substantially improve the model. 

\begin{table}[H]
    \begin{center}
        \begin{tabular}{ |c|c|c| } 
            \hline
            Parameter & Estimate & Standard Error\\
            \hline
            $\alpha$ & 46415.16 & 125.55\\
            $\lambda_1$ & -562.47 & 8.20 \\                        
            $\beta_1$ & 39.39 & 2.18\\
            $\beta_2$ & -0.31 & 0.01\\
            $\gamma_1$ & 125.96 & 9.51\\
            $\phi_{2009}$ & 6712.74 & 86.97\\
            $\phi_{2010}$ & 6904.77 & 87.03\\
            $\phi_{2011}$ & 5798.39 & 85.75\\
            $\phi_{2012}$ & 5325.14 & 87.35\\
            $\phi_{2013}$ & 4601.24 & 85.64\\
            $\phi_{2014}$ & 4662.72 & 85.80\\
            $\phi_{2015}$ & 3530.06 & 85.55\\
            $\phi_{2016}$ & 2647.77 & 86.42\\
            $\phi_{2017}$ & 1839.40 & 86.44\\
            $\phi_{2018}$ & 707.50 & 85.72\\
            $\omega_1$ & -3301.58 & 68.31\\
            $\omega_2$ & 1664.20 & 68.35\\
            $\omega_3$ & 1720.86  & 68.31\\
            $\omega_4$ & 1576.71 & 68.30\\
            $\omega_5$ & 1436.08  & 68.30\\
            $\omega_6$ & -3616.42  & 68.22\\
            \hline
        \end{tabular}
    \end{center}
    \caption{Parameters of the Linear Regression Model for Daily Peak Demand (in MW)}
    \label{tab:OLSResults}
\end{table}

\begin{figure}[htbp]
    \centering
    \includegraphics[width=0.75\linewidth]{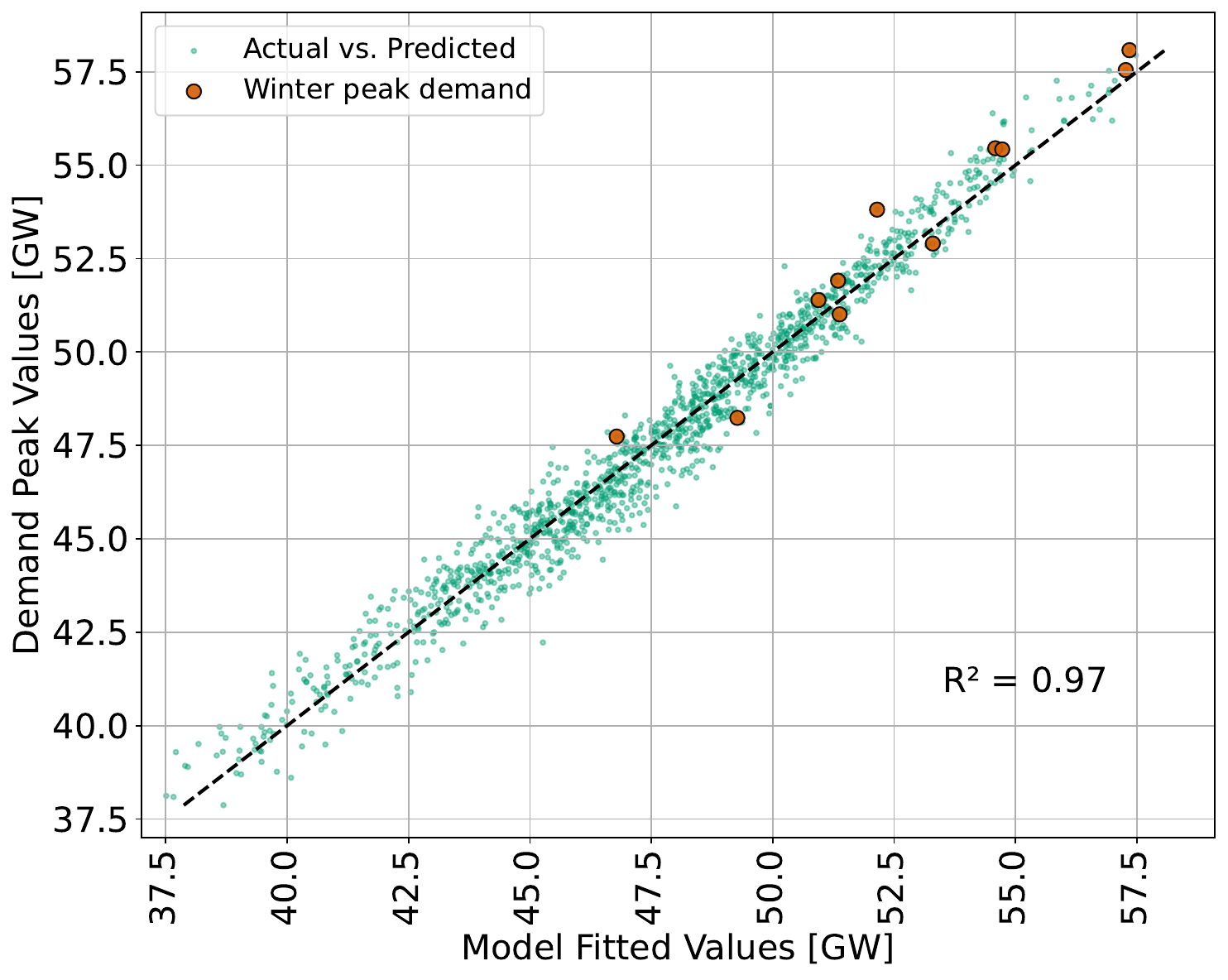}
    \caption{Scatter plot between the model fitted and the daily peak demand values for the subset of data-points between 2009-2019 winter days over the GB. The red dots indicate the peak demand observed in each winter season. See \autoref{sec:assessfit} and \autoref{sec:empres} for consequent discussion of validity of the model.}
    \label{fig:1}
\end{figure}

 \autoref{tab:OLSResults} gives the coefficient estimates and standard errors in the model in MW units. Under the assumption that the residuals are independent over time, all the coefficients are statistically significant. \autoref{fig:1} shows a scatter plot of the model-fitted values and daily peak demand observations. The red dots indicate how well the model estimates the maximum demand during each winter peak compared to the observations. The figure suggests that the model is a good fit to the data overall, with only a few extreme outliers, and an adjusted R-squared value of 0.97. The highest out-turn demands are mainly associated with positive residuals -- however there is no strong evidence from this and other diagnostics that there is any special effect on demand in the coldest conditions, and hence that this observation is other than the inevitable point that the very highest demands will be associated with high values of both the explained component and the residual\footnote{Ongoing work, to be reported elsewhere, suggests that there is  stronger evidence for a special effect on demand (i.e. beyond a smple linear sensitivity to temperature) in the coldest conditions when one models daily aggregate demand (as opposed to daily peak).}.

The quantile-quantile (Q-Q) plot in Figure S1 explores further the behaviour of the residuals, showing that they are close to normally distributed apart from a small deviation in the tails.  As we would expect, there is some positive autocorrelation in the model residuals, with high residuals on one day tending to result in high residuals on the next. Despite this, the coefficient estimates are unbiased, and so the only impact that this autocorrelation has on the fitted model parameters is that the standard errors in \autoref{tab:OLSResults} are underestimates. 
When studying a future system scenario, in addition to the evolution of the coefficients, it is natural to assume that the distribution of residuals will also change from historic values -- thus to account for this possibility, risk calculations for future scenarios are performed using both the empirical historical residuals (equivalent to a standard hindcast calculation) and a stochastic model for residuals. Time correlation in the residuals will not affect calculation results in the `stochastic residuals' model, as only results for expected value indices are presented.

\subsection{Transforming Demand Series to a Target Scenario}

This section presents the approaches used for combining historic weather data with a future scenario of what is connected to the system -- this future scenario  being defined by the values of the coefficients in the regression model.

\subsubsection{Stochastic Residual Model}

The first approach  is to use directly  the future parameter estimates (i.e. of of year effect $\gamma_p$, temperature sensitivity $\lambda_p$ and wind chill $\gamma_p$) associated with a scenario $p$ in the regression model \eqref{DM1}.  It is necessary to add a stochastic residual term to model the part of demand not explained by the regression formula. The resulting model is given by:
\begin{equation}
\begin{aligned}
    \mathrm{D^{STOC}_{p,i,t}} &=  \alpha + \lambda_p [TE]_{i,t}  + \beta_1[DSN]_t  + \beta_2[DSN]_t^2  \\
    &\quad    + \sum_{m=1}^6 \omega_m[DOW]_{m,i,t} + \gamma_p[WS]_{i,t} + \phi_p + \mu_{i,t}\\
    &= \mathrm{\hat{d}_{p,i,t}} + \mu_{i,t}
\end{aligned}
\label{RD1}
\end{equation}
Here, $\phi_p$, $\lambda_p$ and $\gamma_p$ are the coefficients for the year effect, temperature sensitivity, and wind chill, with (for our analysis) values according to the scenarios described in \autoref{tab:Scenarios}. $\mu_{i,t}$ is the normally distributed residual term. As the residual term is a random variable,  $\mathrm{D^{STOC}_{p,i,t}}$ is also a random variable.

\subsubsection{Empirical Residual Model} \label{sec:empres}

The other approach is to map the historical demand series to the future scenario $p$, retaining the empirical residuals from the regression model fit.  This mapping is given by:
\begin{equation}
\begin{aligned}
    \mathrm{d^{EMP}_{p,i,t}} &= \mathrm{d^{EMP}_{i,t}} + (\lambda_p - \lambda_1) [TE]_{i,t} \\ 
    &\quad + (\phi_p -\sum^{2019}_{i=2009} \phi_i[Y]_i)  \\
     &\quad + (\gamma_p - \gamma_1) [WS]_{i,t} \\
\end{aligned}
\label{RD2}
\end{equation}
This mapping simply subtracts off the relevant terms with the fitted values from the historic demand, replacing them with equivalents based on scenario $p$. There is no explicit residual term, as the demand (mapped to scenario $p$) implicitly retains the empirical residual value from the regression fit -- unlike $\mathrm{D^{STOC}_{p,i,t}}$, $\mathrm{d^{EMP}_{p,i,t}}$ is not a random variable.

The `empirical residual' model of this paper follows the same general idea as standard 'hindcast' approaches, in which the demand series is appropriately rescaled to a future scenario, though the detailed implementation may differ. It is often thought that a benefit of using empirical residuals is that these consider what actually happened historically. This is indeed (by definition) a feature of using the empricial historical residuals, but it is actually not necessarily a benefit -- the question is whether in the same circumstances (of date, weather etc), the demand might have been different, through a different value of the residual being realised. In circumstances of limited data on extremes, there is no definitive answer to this without detailed historical investigation of cause and effect in the historical circumstances, but (as mentioned above) based on analysis of the residuals there is no strong evidence that the stochastic residual model is inappropriate --  thus using the empirical residuals might actually not use all of the available information on what demand could have resulted from the historic weather, in addition to what actually did happen. Determining whether a special effect on demand in most severe periods (i.e. beyond the regression terms that apply at more normal temperatures) is required would be an interesting subject for further investigation. However, it is the weather shift approach and not demand modelling per se which is the main topic of this paper, and thus we do not attempt a definitive resolution to this question, as the primary conclusions of the paper apply equally to the empirical and stochastic residual variants of the model.

\subsection{Shifting Model for Generating more Demand Traces}

It is evident from the fitted regression model that for given weather in GB, demand depends strongly on the associated day of week, due to different behavioural patterns \cite{PULLINGER2024122683}; it is likely that similar will apply in most systems. In GB there is also an extended period of around 2 weeks over Christmas\footnote{We do not attempt detailed modelling of the Christmas effect in GB, as the detail of this depends on the day of week on which Christmas falls, which in turn affects the dates between which holiday tends to be taken.} when demand is suppressed due to holidays; an equivalent will be seen in other systems where there is a major holiday period during the peak demand season, but would probably not (for instance) apply in a summer-peaking northern hemisphere system where Christmas is a major holiday (as Christmas would not then be in the peak season). 

As a result, if one analyses historic demand data without accounting for these weekend and Christmas effects, this may result in omission from the analysis of relevant information on severe weather days. Equivalently, one can make much more effective use of the available data by considering all possible alignments of day-of-week with weather. It is also useful to consider the combination of historic weather from the Christmas period, with dates outside Christmas where demand is not suppressed in the same way; greater caution is required here, as this might require longer shifts of weather with respect to date, bringing some (but not necessarily critical) concern about physical realism.

This section describes the model developed for changing the alignment of day-of-week with date, and for shifting weather with respect to date.

\subsubsection{Shifting Association of Day of the Week With Dates}
The first shift model maps the historic data on to any alignment of day-of-week with date. Except for the possibility of shifting weather in and out of the Christmas period, this is very closely equivalent to shifting weather by up to $\pm 3$ days, as in the middle of the peak system the variation of the time-of-year effect with small date changes is very small. As discussed above, there is unlikely to be a direct causal or indirect association between day-of-week and what the weather might be, and so there are no concerns about realism (beyond of course the statistical model needing to be sufficient for the purpose). The model simply swaps out the day-of-week effect associated with the historical date, and swaps in the day-of-week effect associated with the chosen allocation of day-of-week to dates:
\begin{equation}
\begin{aligned}
    \mathrm{^{0,k}d^{EMP}_{p,i,t}} &= \mathrm{d^{EMP}_{p,i,t}}  + \sum_m[DOW]_{m,i,t}\left(\omega_{m+k}-\omega_m \right)\\
\end{aligned}
\label{Shift1}
\end{equation}
where $\mathrm{^{\tau,k}d^{EMP}_{p,i,t}}$ is the adjusted demand with weather shift of $\tau$ days and day-of-week shift $k$ (which without loss of generality we take to be in the range -3 to +3, given that a shift of $k$ is equivalent to a shift of $k \pm 7$). 

\subsubsection{Shifting Weather to Generate Short or Long-Term Variation in Demand}
The second shift model shifts weather relative to date. Similar to the first model, it simply swaps out the weather effects associated with the actual historical weather, and swaps in those associated with the relevant day:
\begin{equation}
\begin{aligned}
    \mathrm{^{\tau,0}d^{EMP}_{p,i,t}} &= \mathrm{d^{EMP}_{p,i,t}} + \lambda_p ([TE]_{i,(t+\tau)} - [TE]_{i,t})  + \gamma_p ([WS]_{i,(t+\tau)} - [WS]_{i,t}) \\
\end{aligned}
\label{Shift2}
\end{equation}
where  $\tau$ is the length of the weather shift - there is no particular limit on the range of $\tau$, but (as discussed earlier) too large a $\tau$ may shift weather by too many days to maintain physical realism. 

\subsubsection{Shift Formulae for Stochastic Residuals}

In principle, it is only the empirical residual model for which these shift formulae are required, as for stochastic residuals one can simply use the regression formula directly. However for computational reasons (which are explained later) it is helpful to define shift formula for $\mathrm{\hat{d}_{i,t}}$ in analogy to that for the empirical residual case of \eqref{Shift1} and \eqref{Shift2}.

\subsection{Examples of Shifting Weather and Day-of-Week}

\begin{figure}[H]
    \centering
    \includegraphics[width=\linewidth, height = 0.7\linewidth]{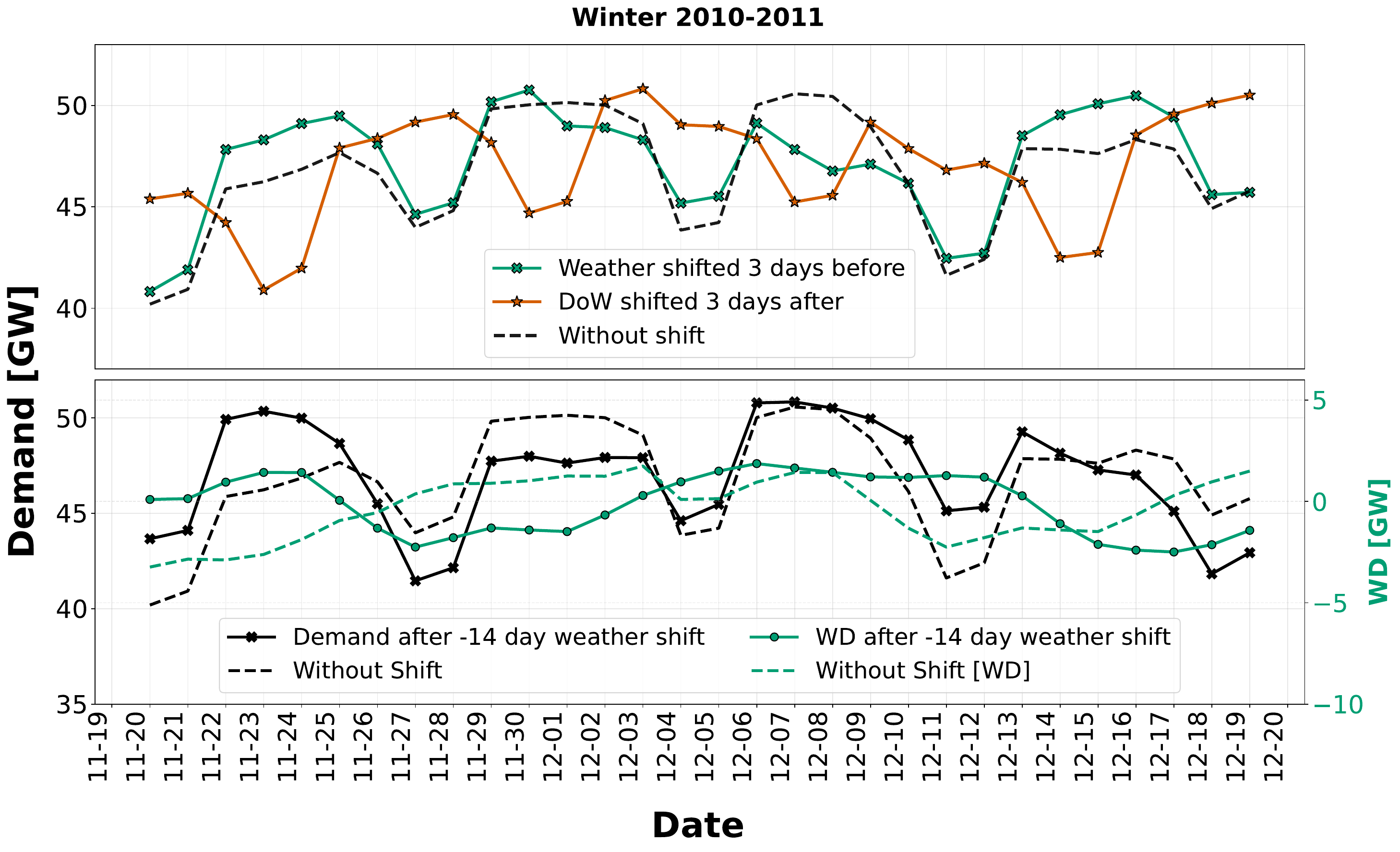}
    
    \caption {Illustration of (a) plus 3-day shift in $[DOW]_{m,i,t}$ and, minus 3-day shift in weather for Nov-Dec 2010 and (b) minus 14-day shift in weather variables to generate the synthetic demand series. The shifted demand series and the WD (includes the component of temperature and wind alone) series are shown in black and green lines, respectively, along with the dotted lines representing their original sequence before the shift. }
    \label{fig:2}
\end{figure}

\autoref{fig:2} presents examples of applying the shift formulae. Panel (a) shows how the daily peak demand changes with 3 day shifts of both the day of the week (DoW) and the weather assignments to date for Nov-Dec 2010. These examples are chosen to show how (due to the weak dependence of the demand on time of year in central winter, and except for when weather is moved in and out of Christmas) the same alignment of day of week with weather can be obtained by shifting either DoW association with date, or the weather series with respect to date.
The two shifted series are identical for risk modelling purposes, unless the empirical residual model is used, in which case the residuals would be associated with date (not DoW or weather).

This also demonstrates how changing the association of day-of-week with weather can give a materially different demand pattern. When the weather driving the high demand period of Monday 6 - Wednesday 8 December is mapped to days-of-week Friday-Monday, the demand on the first day is slightly suppressed, and that on the second two is strongly suppressed due to the weekend effect on demand. It will be seen later that this suppression of demand at weekends reduces the risk of shortfalls at weekends to a minimal level, and removes from the risk assessment the information on weather associated with these days.


The lower panel in \autoref{fig:2} demonstrates a larger shift in weather relative to date, using the model of \eqref{Shift2}. Here the weather is shifted 14 days forward, and a cold period becomes relevant which would otherwise have occurred over Christmas (when demand is suppressed). This is seen in the shifted series from about 5 to 13 December, and in outcome demand (as opposed to the weather component) actually leads to high demand from 6 to 10 December and on the 13th due to weekend effects on other days. Similarly to the demonstration of smaller shifts in the upper panel of the figure, which illustrates how weather occurring at weekends can be brought into assessments using this paper's methods, the lower panel illustrates the potential importance of considering weather over the holiday period. 


\section{Modeling Risk of Shortfalls using Shifted Demand and Generation Series} 
\label{sec:LOLE}

\subsection{Estimating LOLE for Weather and Day-of-Week Shifts} 

The LOLE calculation using empirical residuals is simply the standard hindcast formula using the appropriate demand and wind power time series:
\begin{equation}
        \mathrm{LOLE^{EMP}_{p,i,\tau,k}} = \sum_{t \in T_i} \mathbb{P}\left( X_{i,t} < \mathrm{^{\tau,k}d^{EMP}_{p,i,t}} - [w]_{p,i,t+\tau} \right)
\label{LOL1}
\end{equation}
where $\mathrm{LOLE^{EMP}_{p,i,\tau,k}}$ is the LOLE for system scenario $p$, and weather data from year $i$ with weather and day of week shifts $(\tau, k)$; $T_i$ is  the set of times in year $i$.

With stochastic residuals, the formula used is
\begin{equation}
        \mathrm{LOLE^{STOC}_{p,i,\tau,k}} = \sum_{t \in T_i} \mathbb{P}\left( X_{i,t}+\mu_{i,t} < \mathrm{^{\tau,k}\hat{d}_{p,i,t}} - [w]_{p,i,t+\tau} \right)
\label{LOL2}
\end{equation}
While in principle for stochastic residuals one does not need the shift formula and can just construct the distribution of demand directly from the regression formula, in practice the computational burden is lower if the distributions of the residuals and of available conventional capacity are convoluted once, and the same procedure as for empirical residuals is then followed except in \eqref{LOL1} replacing $X_{t}$ with $X_{t}+\mu_t$, and $\mathrm{^{\tau,k}d^{EMP}_{p,i,t}}$ with $\mathrm{^{\tau,k}\hat{d}_{p,i,t}}$.

To give a controlled experiment, when considering different scenarios of temperature and wind speed sensitivity of demand, installed wind capacity, and of the SD of the residual term, once a scenario of these parameters has been fixed, the year effect on demand is chosen to give a `daily' LOLE of 0.1 days per year. While the GB reliability standard is set at 3 hours per year LOLE, the traditional N American standard of 1 day per 10 years is closer to a level that would actually be deemed acceptable; and because our demand model is for daily peak demand we present the main results for LOLE at time of daily peak demand, with a second set of results presented for hourly LOLE.

\subsection{Effect of DoW Shifts on LOLE}

\begin{figure}[H]
    \centering
    \includegraphics[width=\linewidth, height = 0.7\linewidth]{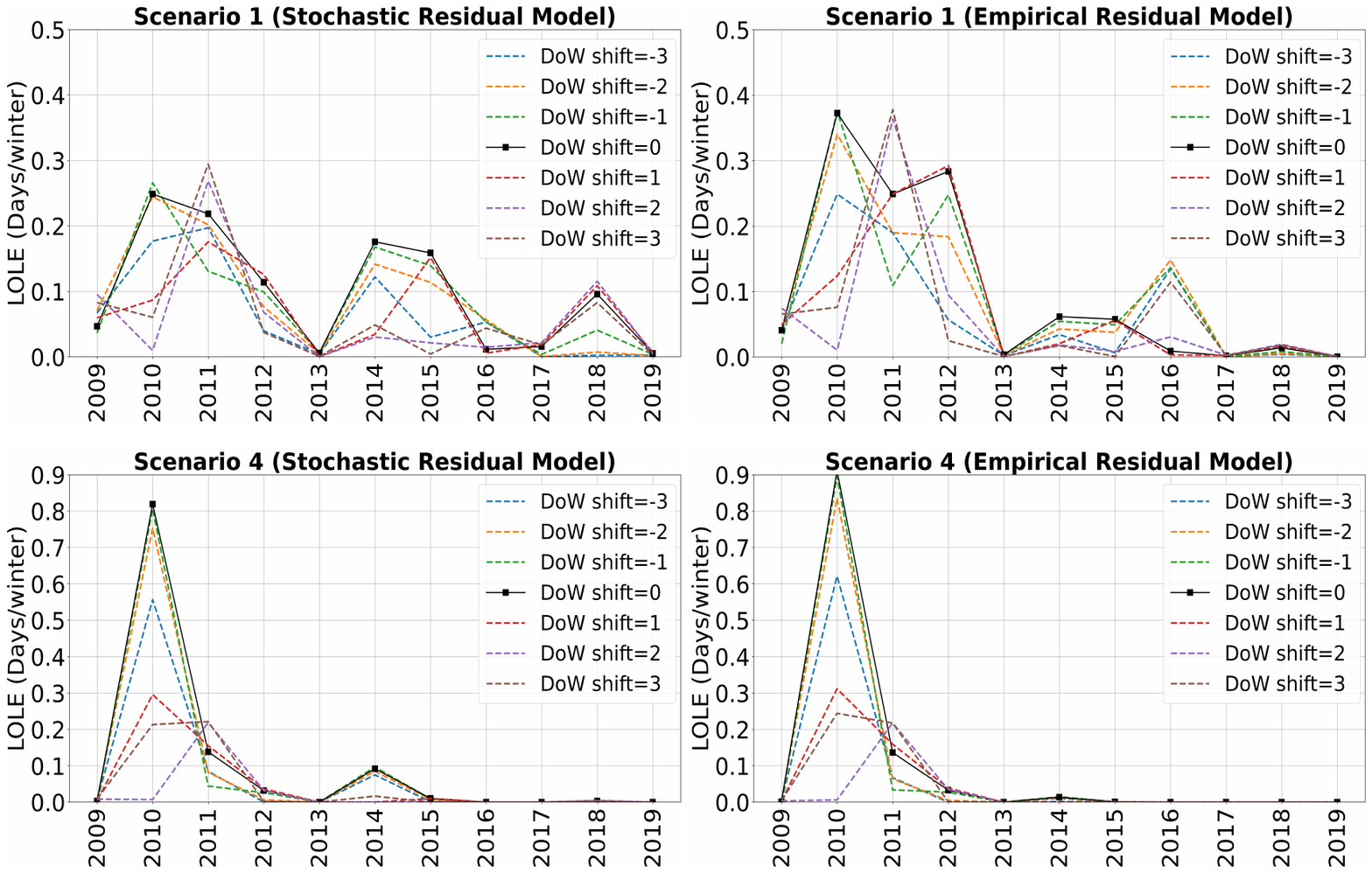}
    \caption{Comparison of LOLE estimates for different winters is shown against different shifts in day of the week assignments with date using the shift formula in \eqref{Shift1}. Fig (a) and (b) compares the energy Scenario 1 (present day) with the stochastic and empirical version of residuals of demand. Same comparison for Scenario 4 (future wind, high heating) is shown in (c) and (d) respectively. }
    \label{fig:3}
\end{figure}

As discussed earlier, the four system scenarios are designed with a range of values of temperature sensitivity and wind capacity, to represent the combination of ongoing growth of wind generation and possible electrification of heating (see \autoref{tab:Scenarios} and relevant text). The black series in \autoref{fig:3} shows the LOLE conditional on each historic year for both Scenario 1 (present day heating and wind capacity) and Scenario 4 (high heating load and wind capacity) with the empirical and stochastic residual models. Equivalent results for Scenarios 2 and 3 are shown in Fig. S2. The more severe years are slightly more dominant in the empirical model than with the stochastic residuals model. However, a much more important effect is the almost total dominance in the highly weather-sensitive Scenario 4 (at least with the historic allocation of DoW to date) by the most severe winter (2010-2011).

Using the shift formulae derived in this paper, it is evident from the other series in \autoref{fig:3} that for a given weather year the estimated risk index can vary very considerably with different DoW assignments. The best illustration of this is Scenario 4 with 2010 weather, where there are only 2 days of week with extreme conditions (in the sense of weather that can drive high demand, with low wind power), which actually are neighbouring days-of-week. Thus there is 1 DoW assignment where these both are weekend days and the LOLE is minimal, 2 DoW assignments with one of these days on a weekend and LOLE of 0.2-0.3 days/winter, and 4 DoW assignments with neither day at a weekend and LOLE above 0.5. This illustrates how the usual practice of only considering one DoW assignment fails to consider the range of possible risk levels conditional on a given weather year, and as a consequence makes very inefficient use of the information contained in the historic data. Note that in all the plots it is only for the actual historical DoW assignment that the LOLE is 0.1 days per year -- the demands are not renormalised for each shift.

\subsection{Effect of Weather Shifts on LOLE}

\begin{figure}[H]
    \centering
    \includegraphics[width=\linewidth, height = 0.7\linewidth]{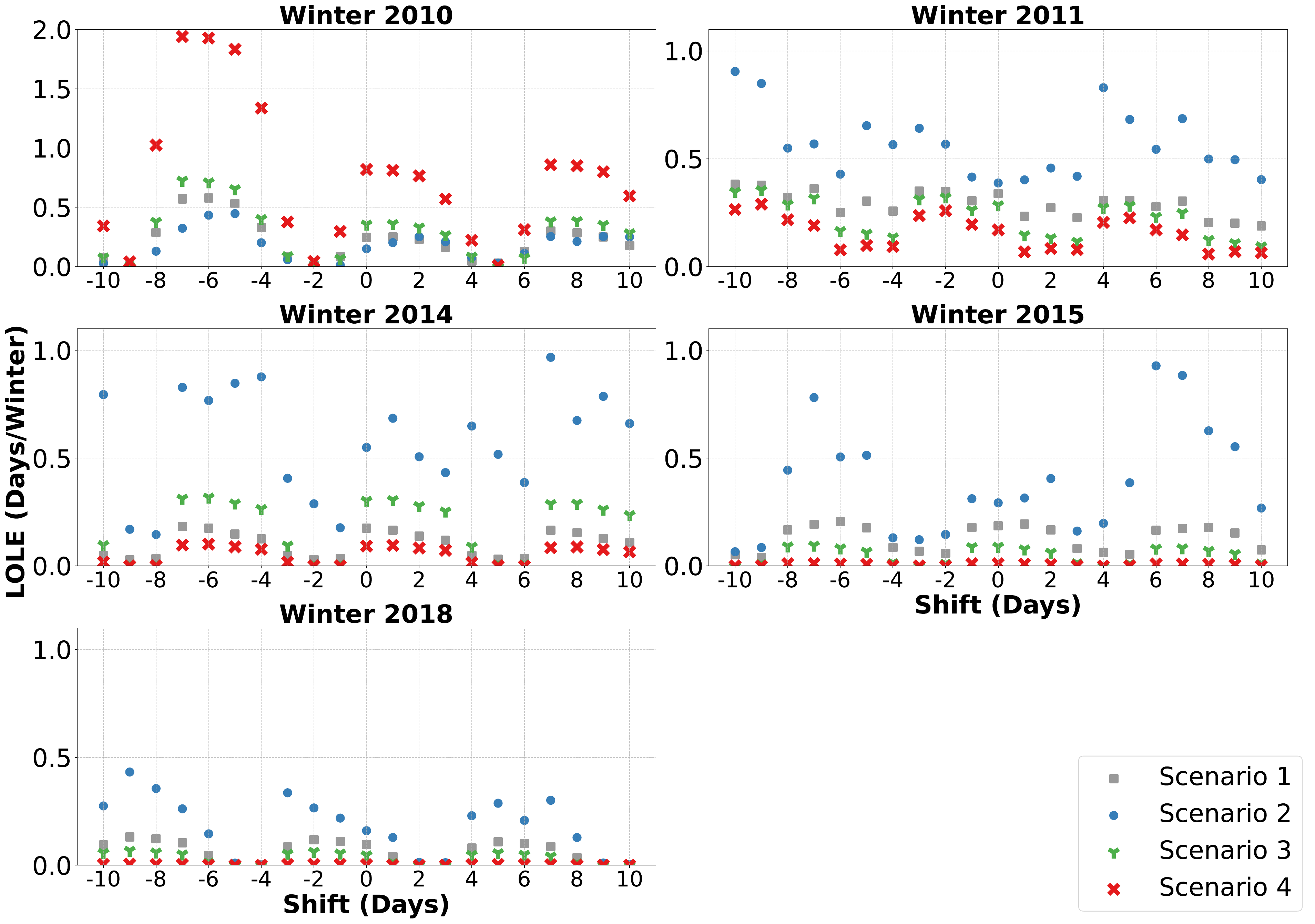}
    \caption{Variation in LOLE using different shifts in weather ($[TE]_{i,t}$ and, $[WS]_{i,t}$) assignments is plotted for each winter in the four scenarios. The three scenarios give the LOLE estimates when  progressive increase in sensitivity of temperature and wind chill coefficients of demand are used against the future wind generation marked in blue, green and red. }
    \label{fig:4}
\end{figure}

\autoref{fig:4} shows the variation of LOLE with shifts of weather relative to date, for all 4 scenarios and a representative selection of 5 winters. Unlike with the DoW shift, where there are only 7 options, there is no one natural limit on the size of the weather shift. This section displays sensitivity analysis for  lengths of shift which in GB would naturally be considered reasonable; we will discuss later the implications for practical risk calculations.

A pattern of period 7 days is clearly seen, arising from the seven combinations of weather and day of week. The plots are however not perfectly periodic -- while for this length of shift, the variation in the time of year effect remains quite small, substantial variations are seen when relevant weather systems move in and out of Christmas. This is seen most clearly for 2010 weather, where there is a very clear weekly pattern but the LOLE is about twice as high for negative shifts which have moved relevant weather out of the Christmas period.

\subsection{Comparison of Daily LOLE and Hourly LOLH Calculations}
\label{hourly}

\begin{figure}[H]
    \centering
    \includegraphics[width=\linewidth, height = 0.35\linewidth]{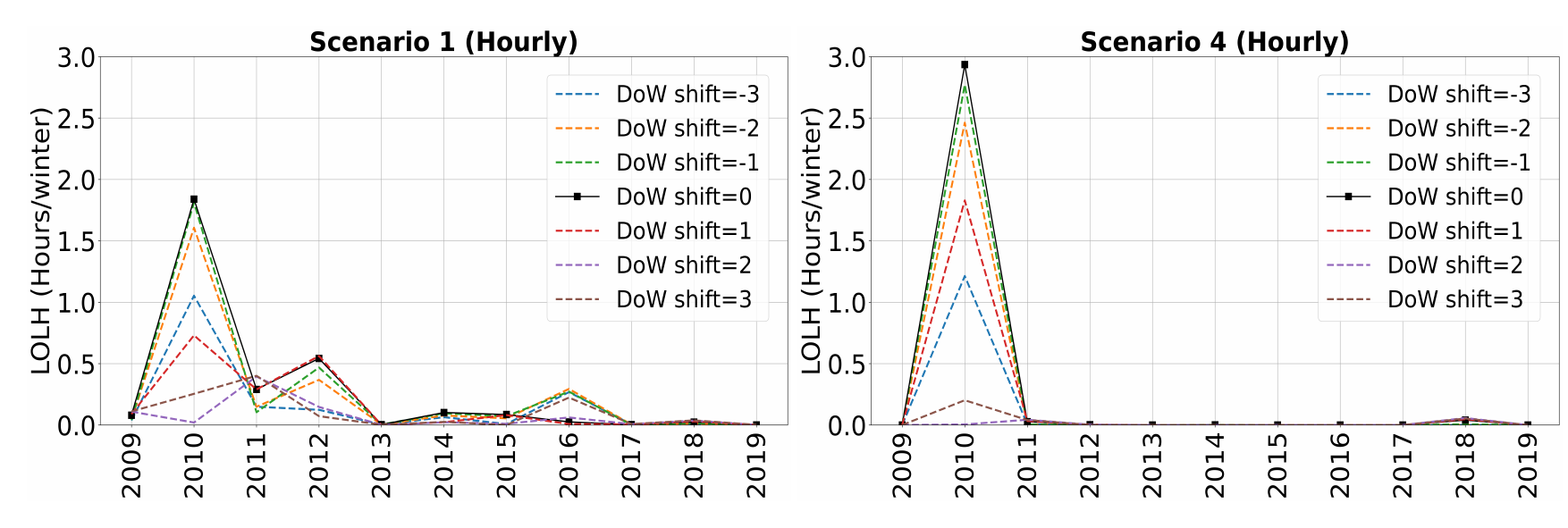}
    \caption{Variation in LOLH for each winter is shown using different shifts in Day of the week assignments using the formula in \eqref{Shift1} for (a) Scenario 1 and (b) Scenario 4. A plus minus 3-day shift in the $[DOW]_{m,i,t}$ is performed on the historical demand series with the empirical residuals using \eqref{Shift1} to account for all the days of the week assignment for a future Scenario.}
    \label{fig:5}
\end{figure}

As discussed above, the main results of this paper are presented for `daily' LOLE, as our statistical demand model is for daily peak demand. However, LOLH is now generally preferred as a resource adequacy metric, and thus \autoref{fig:5} presents the equivalent of the `empirical residuals' model of \autoref{fig:3} for an LOLH calculation. For the demand series, the daily peak demand for a given day is that from the daily peak model; the other hours in the demand series are that day follow the pattern relative to the daily peak seen in the historic data. The same wind series is used as in the daily LOLE calculation, except that all hourly values are used instead of just those from 6pm. The year effect for demand is chosen to given LOLH of 0.3 hours/year averaged across the available historic years.

The broad pattern of the results is  similar between the LOLE and LOLH versions of the plot, though in the LOLH version the dominant weather years are more dominant. However the important point for this paper's conclusions is that conditional on a given historic year, for the LOLH calculation one sees the same wide range of LOLE values as for the daily LOLE calculation. The LOLH version of the weather shift analysis in \autoref{fig:4}  is shown in the supplementary Fig. S3, and again the general pattern of the LOLE and LOLH results are similar with some, though again there are some differences of detail. 

\section{Statistical Interpretation of the LOLE Estimates}
\label{sec:SI of LOLE}

\subsection{Alignment of Weather with Day of Week}

As mentioned above, the alignment of weather with the day of week cycle is, for all practical purposes, arbitrary. It is of course possible for any date to be associated with any day of the week, and in GB (and probably any other power system) shifting weather in time by up to $\pm 3$ days will give a realistic weather time series. The possible consequences of a given weather year have thus not been properly explored unless all seven alignments of weather with day-of-week have been assessed. 

Whether this is done by changing the assignment of day-of-week to date, or by shifting the weather time series by up to 3 days, depends on the application. If a risk assessment is being done for a particular future year then the latter might be natural; if a study is performed for a more generic scenario not tied to a particular year, the former might be preferred. The only substantial difference between the two will occur if as well as shifting weather into and out of weekends, moving weather relative to date also shifts important weather in or out of the Christmas period or winter season. 

There is no complication with the statistical interpretation of this. It can either be interpreted as giving the full range of risk levels that might be associated with a given weather year (see for example Figs. \ref{fig:3} and \ref{fig:5}), or by calculating a risk level conditional on no particular alignment of day-of-week and weather (see \autoref{fig:6}). The latter makes the natural assumption that all 7 day-of-week alignments with weather are equiprobable.

\subsection{Longer Shifts of Weather With Respect to Date}

When shifting the weather time series rather than the day-of-week assignment, it is  possible to perform longer shifts of weather with respect to date. The purpose here would be to move interesting weather systems from during the Christmas period to outside that period, thus removing the Christmas suppression of demand and allowing consideration of the security of supply implications of that real historic weather. One must be careful however, as if the date on which weather occurs is shifted too far then there may be concerns about realism. While there is no natural hard limit on the size of weather shift, and a shift of two weeks (as required to move weather out of the Chistmas holiday) is reasonable in GB, longer shifts of 3-4 weeks are likely to be unrealistic.

It is certainly possible to perform scenario studies, asking what the power system consequences would be if a certain weather system occurred at a different time of year, without giving a statistical interpretation. One could also within this scenario study consider all possible day-of-week assignments to date or weather, as in the previous section. However, it is difficult to assign a statistical interpretation to such a study as there is no natural limit on the realistic size of shift.

\begin{figure}[H]
    \centering  
    \includegraphics[width=\linewidth, height = 0.8\linewidth]{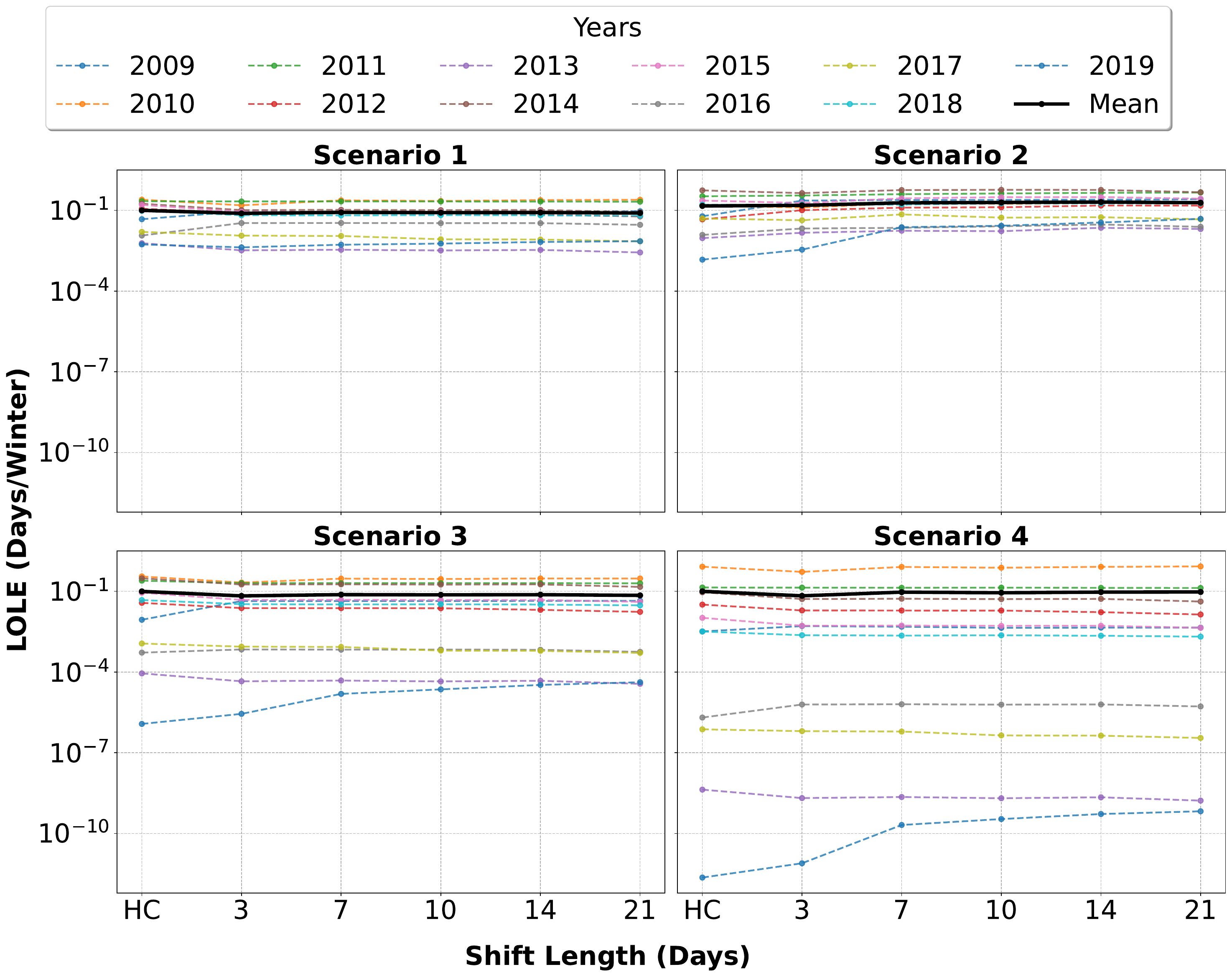}
    \caption{Mean of the LOLE estimate with increasing shift length for each individual winters is shown as a comparison against the Hindcast Estimate (HC). LOLE values on the y-axis are shown on a logarithmic scale. The black line indicates the mean across all the winter years and is used as a check for stability. }
    \label{fig:6}
\end{figure}

An alternative approach is to calculate the LOLE conditional on a given maximum weather shift. The difficulty here is that computing this conditional LOLE requires an assessment of the relative likelihoods of the possible weather shifts, up to this maximum. One possible approach is to assume that all weather shifts are equally likely. This would seem to be a reasonable assumption, if the maximum shift is chosen such that there is no reason to believe that the weather-date assignment for the most extreme shift would be any less likely to occur in practice than the historical weather-date assignment. If all weather shifts up to the maximum shift are equally likely, then the LOLE conditional on a given maximum weather shift can be calculated as the mean across the LOLE values, conditional on the possible weather shifts within that window. \autoref{fig:6} shows these LOLEs for hindcast, and $\pm 3$, $(-7, +6)$, $\pm 10$, $(-14, +13)$ and $(-21, +20)$ maximum weather shifts. These results are displayed conditional both on each weather year, and averaged over all weather years. If a given result is stable as  the maximum shift length is increased up to some credible maximum, then one can be confident in that as the risk level conditional on the weather in the given year(s). On the other hand, if the result is not stable, then one has to accept that one does not have such a confident estimate for the LOLE conditional on the given weather sequence. As we expect seasonal trends in weather as we move from winter to spring, it is important to pre-specify a credible maximum when checking the stability of these results. If this maximum is too high then stability will never be achieved.


\autoref{fig:6} shows, as one would expect, that the LOLE values averaged across all years are more stable with maximum shift length than some of those conditional on a given weather year. This is both because of averaging reducing the influence of individual years taken in isolation, and because of some cancellation between the changes with maximum shift length of the LOLE values conditional on different weather years (most notably in Scenario 2). As per previous sections, where there is substantial dependence of LOLE on the maximum shift length, this is due to significant weather moving in or out of Christmas.

\section{Conclusion}

\label{sec:conc}



This paper has demonstrated an approach for shifting weather or day-of-week (DoW) relative to date in demand modelling. The key contribution of this paper is the consequence for risk modelling of considering all the seven  possible alignments of DoW and weather, and resulting recommendations for practical resource adequacy calculations.

For predictive modelling purposes, the assignment of DoW to dates may have some relevance if the risk calculation is for a particular future year; the examples given here have been for generic scenarios not tied to a particular future year, in which case there is no preferred assignment of DoWs to date. More importantly, even if for a particular study there is reason to use a given assignment of DoW to date, in most systems (and certainly in GB) any single assignment of days-of-week to a weather time series will be entirely arbitrary -- it will be very rare to find a region with a meaningful difference in the climate for dates 3 days apart.

As a consequence, if all 7 alignments of DoW with weather have not been included in an assessment (through DoW assignment to dates, or through shifting weather with respect to data, as appropriate), then the consequences of a given weather series for RA risk cannot be fully understood -- if there is a substantial variation of demand level with DoW, then risk calculation results might depend strongly and arbitrarily on the alignment of DoW and weather considered.

This is certainly the case in GB, where demand is strongly suppressed at weekends. For instance the most challenging weather days (i.e. highest values of demand minus wind in systems scenarios considered here)  in the  time series for 2010-11 could all appear on weekdays (making this the most severe winter in the dataset used), or all at weekends (reducing  the risk of shortfalls to a very low level in the calculations performed). Considering the 7 assignments of day of week with weather (including through shifting weather wish respect to date) presents no difficulties with statistical interpretation -- one can either average results over the 7 possibilities and interpret this as an improved estimate of an expected value index conditional on the weather year; or interpret the 7 values as providing a range of risk levels conditional on the combination of weather year and DoW alignment. 

Longer shifts of weather with respect to date may also be of interest, precisely because (in GB context) this allows the `unmasking' of interesting weather which happens at Christmas, or conversely consideration of how the severity of a winter would change if severe conditions were moved to the Christmas period. Similar would apply in any system where there is an extended holiday period in the peak demand season, for instance Christmas in much of Europe, but not in all of N America where some regions are summer-peaking. One pragmatic approach demonstrated here is to consider sensitivity of calculation results to the maximum shift considered, if an overall risk index is obtained by averaging over results from the range of shifts considered; if the result is then stable over a range of maximum shifts then this provides a stable result conditional on the weather year. Another approach would be to take a  scenario-driven approach, and identify the most severe alignment of the weather time series with date; this provides an upper bound on the severity of the  given weather year.

\section{CRediT authorship contribution statement}

\textbf{Aninda Bhattacharya} - Writing - Original Draft, Methodology, Formal Analysis, Visualization, Validation. \textbf{Chris Dent} - Writing - Review and Editing, Conceptualization, Methodology, Supervision. \textbf{Amy L. Wilson} - Writing - Review and Editing, Conceptualization, Methodology, Supervision. \textbf{Gabriele C. Hegerl} - Writing - Review and Editing; Methodology, Supervision.

\section{Declaration of competing interest}

The authors declare that there are no known competing financial interests or personal relationships that have influenced the work in the paper.

\section{Acknowledgement}

The authors acknowledge support from the University of Edinburgh, UK and the National Environmental Research Council (NERC) under UK Research and Innovation (UKRI) to carry out this project. The authors also thank the National Energy System Operator (NESO) for providing electricity demand and generation data, and associated discussions regarding the work; and Hannah Bloomfield (Newcastle University) for provision of weather data. They further acknowledge productive discussions with members of the IEEE PES Resource Adequacy Working Group,  with colleagues at the NERC Probabilistic Analysis forum, and with Estelle McCool (whose undergraduate dissertation developed an early version of the shift model used here).  

\newpage
\bibliographystyle{elsarticle-num}
\bibliography{ref}

\end{document}